# Approaching the Low Optical Loss Limit of Plasmonics using Potassium Metals

Yuhan Yang[1]†, Yi Zhang[2]†, Yuhong Shi[1]†, PengCheng Yao[1]†, Hanyu Fu[1], Jacob B. Khurgin[3]*, Fengrui Hu[1], Jia Zhu[1], Shining Zhu[1], Lin Zhou[1]*

[1]National Laboratory of Solid State Microstructures, College of Engineering and Applied Sciences, Key Laboratory of Intelligent Optical Sensing and Manipulation, Ministry of Education, Nanjing University, Nanjing, P. R. China, 210093

[2]School of Information and Electronic Engineering (Sussex Artificial Intelligence Institute), Zhejiang Gongshang University, Hangzhou 310018, People's Republic of China.

[3]Dept of Electrical and Computer Engineering, Whiting School of Engineering

Johns Hopkins University, 314 Barton Hall, 3400 N. Charles St., Baltimore, MD 21218

†These authors contributed equally to this work.

*Email: linzhou@nju.edu.cn, jakek@jhu.edu

Plasmonics enables the miniaturization of photonic devices beyond the optical diffraction limit, yet its potential is hindered by inherently large ohmic losses. Hence, it is prudent to explore low-loss alternatives to the current mainstay of plasmonics—the noble metals. In this work, we demonstrate the potential of potassium as a plasmonic material with intrinsically low losses in the optical region. The ultra-flat, high-quality potassium film, fabricated via a rapid slipping-assisted oxide-free crystallization process, achieves an experimentally observed optical damping rate of 3.7 meV, with a measured imaginary permittivity of approximately 0.1 across the entire visible to near-infrared range (400 – 2000 nm). Near-field optical spectroscopic measurements further confirmed the reduced losses by revealing deeply subwavelength confinement of optical modes. This result enhances our understanding of the factors governing plasmonic materials and devices and establishes a new platform for exploring extreme light–matter interactions in a variety of plasmonic systems.

Because of its distinctive power to manipulate light fields beyond the optical diffraction limit, plasmonics has been pursued for enhanced light-matter interactions[1–4], super-resolution imaging[5–7], low-threshold nanolasing[8–12], etc. This unique capability of plasmonics, however, is hindered by their inherently high optical losses ever present even in the noble metals that are currently the most widely used plasmonic materials. Alternative low-loss plasmonic materials has always been the focus of attention in the recent years [13–15].

The past decade has witnessed development of various innovative fabrication processes such as self-assembling[16], templated stripping[17] and epitaxial growth technique[18,19] aimed to improve the quality of metal films and minimize optical losses, but still even in silver the losses are too high to satisfy requirements of most plasmonic devices other than sensors. Significant efforts have also been devoted to the non-noble metal such as borophene[20], hyper-gap organic conductors[21] and sodium[22], in both experimental and theoretical studies. Even though a couple of improved thermal assisted techniques have been reported for nanostructured sodium metals and/or their alloys[23-25], all of which focused on the low loss property of metasurface instead of unstructured alkali metals that are mainly stemmed from hybrid plasmon effects[26]. In addition, as ideal low-loss plasmonic metals at non-vacuum conditions are chemically reactive, the reported fabrication processes are still far from precise control and thus suffering from poor material quality and large structure heterogeneity[22-25], making the low loss limit achievable in plasmonic metals (especially in even more chemical reactive alternatives) still illusive thus far. The underlying significant obstacles are basically stemming not only from the fabrication challenges of these highly reactive materials[22-26] but also from the incomplete understanding of the scattering mechanisms that govern optical losses.

In this work, we theoretically predict potassium (K) to be the ideal low-loss plasmonic metal in optical regime. This conclusion is based on an optical-loss phase diagram developed using the decomposed normal (N) and umklapp (U) electron–phonon scattering damping rates. In the experiment, a unified ultrafast slipping-assisted oxide-free crystallization process is developed to overcome the high chemical reactivity of K and successfully fabricate near-single-crystalline potassium films in wafer scale. The prepared high-quality potassium sample is shown to reduce the optical damping of best silver metals by 1- 2 orders of magnitudes thus enabling deep subwavelength field confinement in surface plasmon polariton (SPP) configuration. It thus establishes a new pathway to brake the usual trade-off between low optical losses and strong confinement.

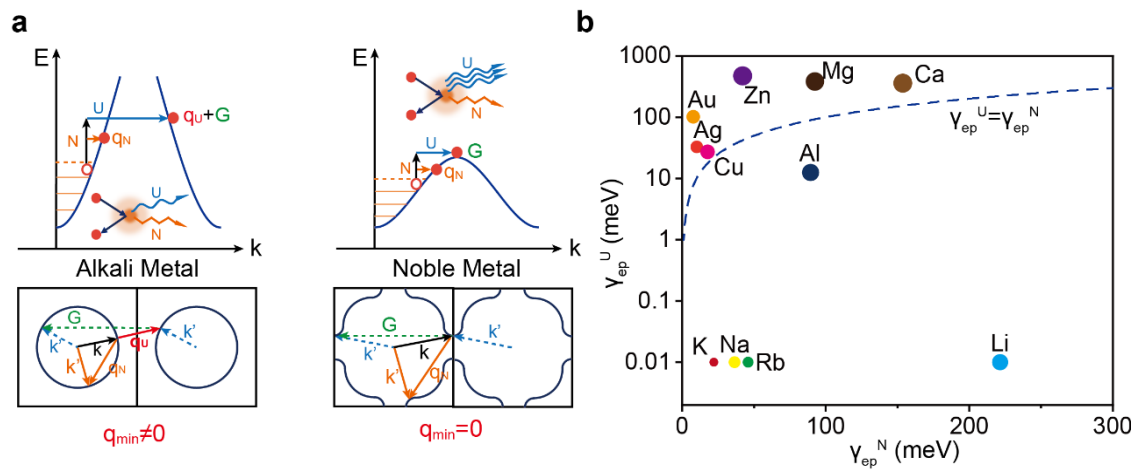

**Figure 1. Theoretical prediction on record low loss potassium metals.** (a) The qualitative band structure based optical transition comparisons and Fermi surface profiles of alkali metals (left panel) and noble metals (right panel) in terms of the decomposed N and U scattering processes (here $k$ and $k'$ are initial and final states, $q_N$ and $q_U$ are phonon involved in N and U scattering, respectively, $G$ is reciprocal lattice vector.). (b) The $(\gamma_{ep}{}^N, \gamma_{ep}{}^U)$ resolved optical loss phase diagram among versatile plasmonic metals in terms of the quantitatively calculated U and N scattering enabled e-p scattering rate at room temperature ($T$ = 298 K). Here the size of circles represents the magnitude of the total damping rate.

We first of all systematically consider the intrinsic optical losses of plasmonic metals primarily originating from the intraband optical transitions out of the intrinsic interband absorption window. The plasmonic damping in optical regime are well known as the sum of three damping rates of $\gamma_{e\text{-}p}$, $\gamma_{e\text{-}e}$ and $\gamma_{e\text{-}s}$ with respect to the electron-phonon (*e-p*), electron-electron (*e-e*), and electron-surface scatterings (*e-s*), respectively, reading as $\gamma_{intra} = \gamma_{e-p} + \gamma_{e-e} + \gamma_{e-s}$. The last contributor to damping—also called Landau damping or Kreibig damping[27–29]—is inversely proportional to the Fermi velocity but mainly dictated by geometry (surface-to-mode-volume ratio). Despite the fact that the lower Fermi velocity in alkali metals leads to reduced surface scattering, this benefit is not significant enough to explain the large loss difference when compared to noble metals. Hence only the first two terms ($\gamma_{e\text{-}p}$, $\gamma_{e\text{-}e}$), determined by the intricacies of Fermi surface profiles and electronic band structures nearby Fermi level $E_F$, are considered below. More specifically, as scattering enabled absorption probability matters the most, the first two terms are highly dependent on the density of states of free electrons above $E_F$ as well as the minimum distance $q_{min}$ between Fermi surfaces in adjacent Brillouin zone (BZ), the latter of which directly determines the proportion of U scattering initiated absorption. Therefore, to minimize the U scattering-based optical transition and achieve the lowest possible loss, the more promising plasmonic candidates among various metals must possess an ideal parabolic band structure and a nearly spherical Fermi surface, characteristics exemplified by alkali metals (as schematically illustrated in Fig. 1a). While the Fermi surface in alkali metals remains far from the BZ boundary, leading to a low probability of U processes, the situation differs in noble metals (right panel of Fig. 1a). There, the Fermi surface intersects the BZ boundary, which is what allows U processes to occur. We start with the electron-phonon damping term $\gamma_{e-p}$ which can be decomposed into two sub-terms ($\gamma_{ep}^N, \gamma_{ep}^U$) independently enabled by N and U scatterings, respectively, $\gamma_{ep} = \gamma_{ep}^N + \gamma_{ep}^U$. At room temperature, the $\gamma_{ep}^N$ term can be evaluated based on the 1st and 2nd plane-wave model in terms of pseudopotential theory[30,31], as expressed by $\gamma_{ep}^N = (4\pi E_F^2 n_F k_B T)/18\hbar\rho v_s^2$ while $\gamma_{ep}^U$ is positively proportional to $\gamma_{ep}^N$ given by $\gamma_{ep}^U = \alpha\gamma_{ep}^N$ (See Supporting information, Note S. I. 1 for more details), where $n_F$ is the state density of electron at Fermi surface, $k_B$ is the Boltzmann constant, $\rho$ is the mass density, $v_s$ is the speed of longitudinal phonons of the targeting metals, $\alpha$ is a dimensionless quantity that characterizes the magnitude of U scattering.

More quantitative calculation results of the ($\gamma_{ep}^N, \gamma_{ep}^U$) resolved two-dimensional plot of optical damping rates among versatile potential plasmonic metals are depicted in Fig.1(b). One can clearly observe that, because of the extremely low U scattering enabled optical damping, the data points of alkali metals (Na and K) lie in the low *e-p* loss region, while other potential plasmonic metals have relatively higher e-p scattering rate due to large proportion of the U scattering contributions. In addition, potassium exhibits distinctly lower Fermi energy among alkali metals, which means potassium simultaneously possesses a much lower N scattering rate compared to sodium or lithium metals. Thus, potassium fits the definition of the holy grail room-temperature plasmonic metals featured by the lowest *e-p* scattering rate among all metals (with Rb and Cs excluded here, as they cannot exist in solid state at room temperature).

As for electron–electron (e–e) scattering, the rate $\gamma_{e\text{-}e}$ shows an even stronger dependence on the fraction of U processes, since in N scattering events the total electron velocity tends to be conserved, resulting in no net dissipation[32]. Consequently, the e–e scattering rate in alkali metals is significantly lower than in noble metals. A similar trend applies to the electron–surface (e–s) scattering loss term. As previously noted, the combination of a lower Fermi velocity and a larger skin depth in alkali metals leads to an exceptionally low e–s scattering rate compared to other metals (see more details in Supporting Information, Note S. I. 1). Thus, from both qualitative physical insight and quantitative modeling of scattering-induced losses, potassium emerges as a metal with exceptionally low optical dissipation—indeed, capable of outperforming the traditional low-loss limit of plasmonic metals governed by intraband transitions.

To experimentally validate the record-low intrinsic optical damping in potassium, it is essential to fabricate high-quality metal films with high crystallinity and atomically flat surfaces, while minimizing oxidation effects. Unfortunately, due to potassium's well-known extreme chemical reactivity—exceeding even that of sodium[33]—films prepared by either ultra-high-vacuum deposition[34] or micro-injection techniques have consistently suffered from oxidation, leading to poor material quality and/or huge structure heterogeneity. Moreover, the thermo-assisted spin-coating method that proved effective for sodium metals is not suitable for potassium, as it operates far from the conditions required for near-crystalline solidification[22] (see more details in Supporting Information S. II. 1).

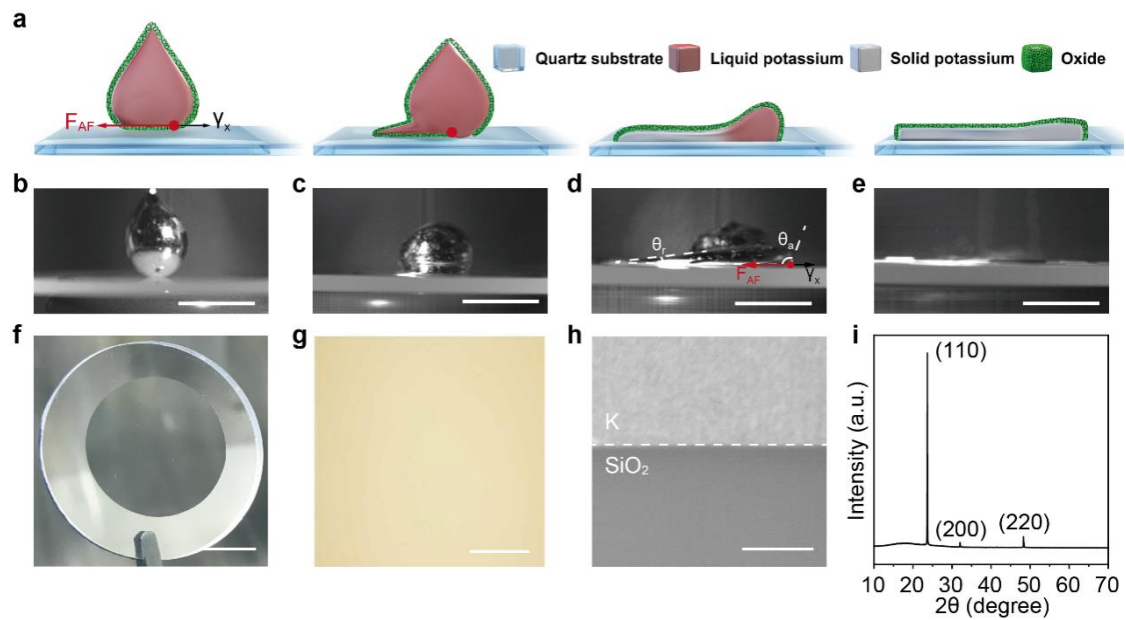

**Figure 2. Fabrication process of SOC-based potassium metal film.** (a) Schematic of the SOC process and (b-e) are the optical images of the corresponding procedure taken by high-speed camera (Scale bar: 5 mm). (f) The large-scale optical photograph of the prepared potassium film (Scale bar: 1 cm). (g) The partial magnification of the light microscope image (Scale bar: 300 μm). (h) FIB cross-section image of the K/quartz film. The dashed line clearly outlines the interface of the upper K film and the underlying quartz substrate, showing a flat morphology without evidence of an oxide layer (Scale bar: 1 μm). (i) XRD test result of prepared K film.

Here by decoupling the oxidation and solidification processes via ultra-fast dicing of oxide-encapsulated hot liquid droplet of potassium metals, we have developed an ultrafast slipping assisted oxide-free crystallization (SOC) process to avoid oxidation and form high-crystal-quality K film. As illustrated in Fig. 2(a), after removing the original oxide and/or impurity shell through vertically aligned quartz tube (Supporting Information S. II. 2) in the inert gas glovebox (widely available non-vacuum conditions), the fresh K droplet undergoes a free-fall and slips against a rapidly moving quartz substrate, solidify in oxide-isolated limited space nearby interface and finally form high-quality metal film (see related optical images in Fig. 2(b-e) taken by the commercial high-speed camera, 5F01C, 1280 × 1024 resolution at 2000 fps). Note that the extremely large tangential-velocity mismatch between potassium liquid and quartz substrate plays crucial roles to ensure the oxide-free limiting-space crystallization, which induces huge lateral adhesion forces ($F_{AF}$) larger than the horizontal component of surface tension ($\gamma_x$), allowing fresh liquid potassium to flow out and directly adhere to the substrate without oxygen exposure. More specifically, when K droplet slides on an ultra-flat stiff surface, the lateral adhesion forces ($F_{AF}$) can be expressed as[35]:

$$F_{AF}(v) = \frac{24}{\pi^3}\gamma D\big(\cos\theta_r(v) - \cos\theta_a(v)\big), \tag{1}$$

where $\gamma$ represents the surface tension coefficient, 106 mN/m[36] for potassium liquid. $D$ denotes the diameter of the contact region. $\theta_a$ and $\theta_r$ stands for the dynamic advancing and receding contact angles, respectively. These two contact angles depend on the relative velocity $v$ of the droplet with respect to substrate: as $v$ increases, $\theta_a$ increases, whereas $\theta_r$ decreases. It thus can be inferred that $F_{AF} \propto v$. In other words, when $v$ is sufficiently high, $F_{AF}$ can exceed the surface tension of the potassium droplet, tearing K droplet's surface shell, thereby allowing the fresh liquid K inside to flow out and directly contact the substrate. Therefore, at the optimized $v$ of 12 m/s, the measured $F_{AF}$ reaches approximately 4.5 mN [Fig. 2(d)], significantly surpassing the droplet's horizontal component of surface tension of 1.1 mN*cos (67°) = 0.43 mN (characteristic droplet size ~ 1 cm), leading to the split of the potassium drop's surface and thus ensuring the instantaneous homogeneous potassium film formation with oxidation effectively excluded. The fresh liquid potassium then solidifies along the path, forming a high-quality K film with an oxide-free interface at the substrate (see more details in Supplementary Movie S1).

More details about the prepared high-quality potassium metals are shown in Fig. 2(f-i). Fig. 2(f) shows the wafer-scale optical photograph of the ultra-smooth potassium film prepared by the SOC method, with the representative magnified optical microscope image depicted in Fig. 2(g). No impurity nor defects are observable over macroscopic areas (> 1 mm$^2$), ensuring the large-scale reliability for both fundamental measurements and device fabrications with respect to conventional methods for highly reactive metals. To further verify the quality of the prepared K /SiO$_2$ interface, the cross-sectional scanning electron microscope (SEM) image is taken by the dual-beam Focused ion beam (FIB) technique (Scios Dual Beam, FEI Company, 30 kV) combined with the self-built vacuum transfer chamber (see details for Methods), as shown in Fig. 1(h). The SOC prepared K/SiO$_2$ surface is ultra-flat without any micro-bubbles or wrinkles profiting from rational wettability control. To further evaluate the crystallization of the prepared potassium film, the standard X-ray diffraction (XRD) characterizations were carried out (test method shows in Supporting Information S. II. 4). As depicted in Fig. 2(i), the SOC enabled potassium film exhibits almost near crystalline nature with the main peak at 110 crystal orientation, which is of much higher quality of potassium compared with pure thermal-prepared sodium in literature (Fig. S12). All these evidences suggest that the proposed SOC method can produce near-oxide-free, large-smooth-area, highly crystalline potassium films up to the centimeter scale in the non-vacuum conditions, laying the unified foundation for facile fabrication of high-activity metal films.

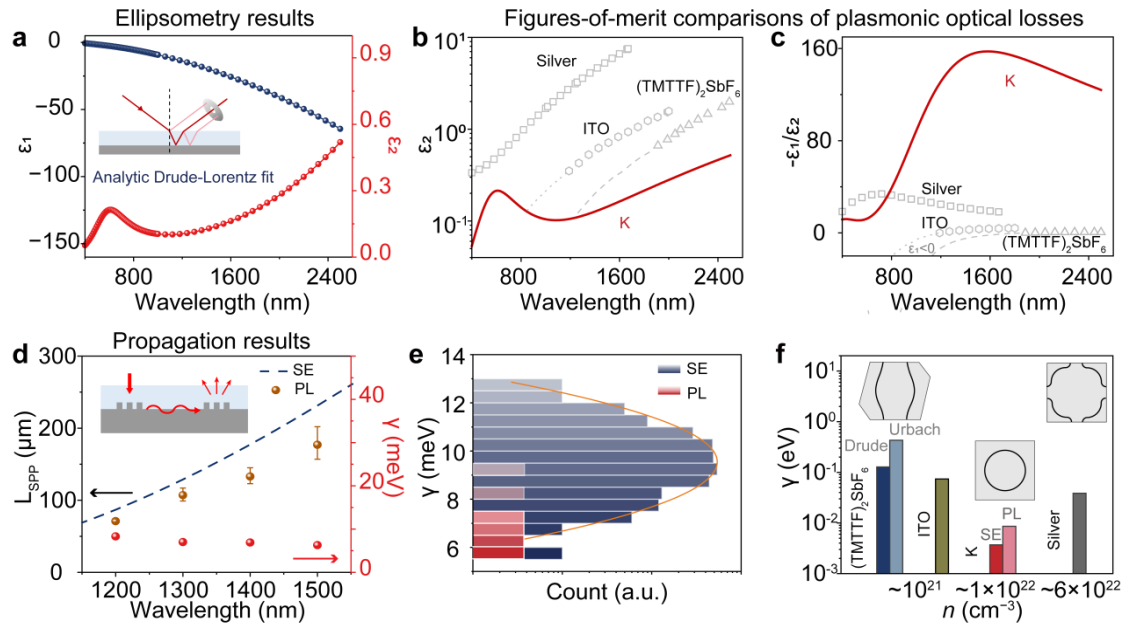

**Figure 3. Optical losses of potassium characterized by far-field spectroscopy.** (a) Real and imaginary parts of the dielectric functions ($\varepsilon=\varepsilon_1+i\varepsilon_2$) of potassium film measured with spectroscopic ellipsometer. The inset illustration shows the schematic diagram of the measurement method. (b)-(c) $\varepsilon_2$ and $-\varepsilon_1/i\varepsilon_2$ comparison with reported low loss plasmonic metals (Ag, ITO and $(TMTTF)_2SbF_6$), respectively. (d) The propagation length and the γ retrieved from propagation length method. The inset illustration shows the measured propagation length at different wavelengths of near-infrared on the potassium–quartz interface. The dotted line is the calculated propagation length based on the dielectric functions of potassium film measured with spectroscopic ellipsometer. (e) The statistical data of γ retrieved from both SE and PL spectroscopic data. (f) Comparions of experimentally reported milestone γ values of mainstream holy-grail plasmonic materials in terms of the free carrier density and Fermi surface profiles.

To analyze quantitatively the intrinsic optical losses of the prepared potassium film, the dielectric functions of about 300 data points mapping at 5 mm*5 mm area in 400 - 1600 nm were measured through commercial spectroscopic ellipsometer (SE) (RC2, J.A. Woollam Corporation). A standard Drude-Lorentz model (without extra nonphysical parameters) work well on fitting SE results, expressed as a sum of intra and interband contributions[37]:

$$\varepsilon(\omega) = \varepsilon_\infty - \frac{\omega_p^2}{\omega^2 + i\omega\gamma} + \frac{f\omega_p^2}{\omega_1^2 - \omega^2 - i\omega\gamma_1}, \quad (2)$$

where $\varepsilon_\infty$ is background permittivity, $\omega_p$ is bulk plasma frequency, $\gamma$ is Drude (momentum) damping rate of free carriers, $f$, $\omega_1$ and $\gamma_1$ are Lorentz amplitude, resonant frequency of interband transition and interband damping rate, respectively. Fig. 3(a) shows a typical fitted dielectric function. The weak absorption peak (~ 600 nm) matches well with the well-known s-to-p interband transition of K (110). For comparison, in the near infrared, the measured $\varepsilon_2$ of SOC-enabled K is merely ~ 1/100 of Ag[16,19] and even ~ 1/10 of transparent conductors such as ITO and

$(TMTTF)_2SbF_6$[21] [Fig. 3(b)]. Fig. 3(c) further demonstrates the comparison of the figure of merit (FOM, defined by $-\varepsilon_1/\varepsilon_2$[38]), indicating that the SOC enabled K have for the first time achieved the state-of-the-art optical performance in the near-infrared range, surpassing other reported low loss metals by one or two orders of magnitudes.

Apart from the measured low optical loss effect validated by the SE data, γ can also be independently retrieved from measured propagation length (PL) of SPP traveling along semi-infinite plasmonic waveguide, serving as an alternative experimental evidence for the low-loss nature of SOC enabled K (see more details in Supporting Information S. II. 6). Various propagation separations were measured to fit the intensity decay curve (Fig. S16), to obtain the propagation lengths ($L_{SPP}$) at different excitation wavelengths on the potassium–quartz interface. Fig. 3(d) also gives the experimental results of $L_{SPP}$ (the magnified error bar is shown in Fig. S15) and the lowest available γ retrieved from the PL data comes to ~ 6.3 meV at 1500 nm (fitting details in Supporting Information S. II. 7). Fig. 3e compares the statistical data of γ extracted using the PL and SE methods, showing consistent wavelength-dependent tendency (lower damping rate for longer excitation wavelength) in literature[22], indicating the agreement with the SE-based record of 3.7 meV (solid line) except for the slight increase probably originating from extra defects and/or impurity during fabrication of the coupling structures. Based on the above independently measured results, a straightforward comparison of $\gamma$ among the best SOC-prepared potassium and mainstream low-loss candidates in literature is depicted in Fig. 3(f) in terms of free carrier density and Fermi surface structures. Our K film wins the record plasmonic material out by reducing all the figures of merit of optical losses ~ 1-2 orders of magnitude compared with the milestone candidates, showing clear correlation with both distinctly reduced carrier density and near-parabolic Fermi surface profiles with minimized U scattering as predicted in Fig. 1.

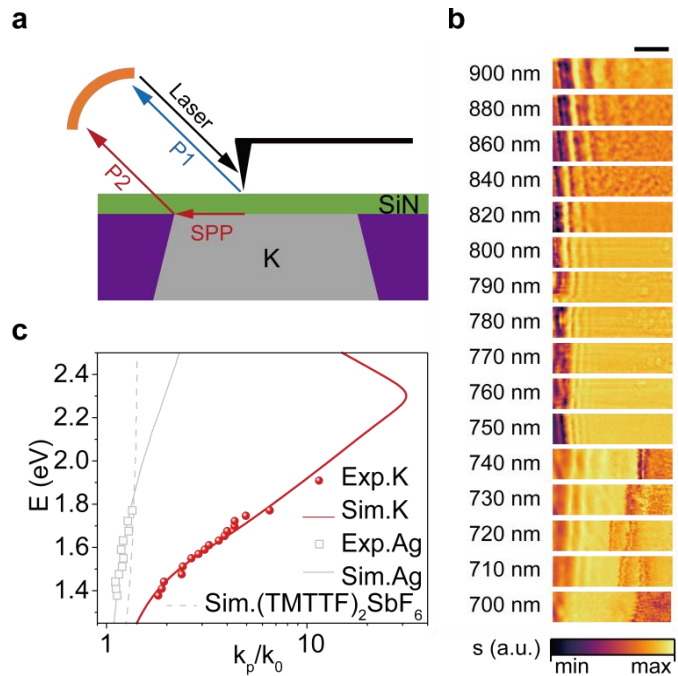

**Figure 4. Low-loss initiated deeply optical confinement of potassium by near-field spectroscopy.** (a) Illustration of s-SNOM experimental set-up to characterize K-SiN-Air SPP device. (b) s-SNOM imaging data of incident light range from 700 - 900 nm. The scale bar for 750 - 900 nm is 1 μm, for 700 - 740 nm is 0.5 μm. (c) Optical confinement comparison of SPP modes excited in metal-SiN-Air device for K, Ag and bulk mode for (TMTTF)$_2$SbF$_6$ both in theoretical (line) and experimental (dotted) results.

Next, to demonstrate how reduced loss allows one to achieve strong optical confinement in SPP mode, tip-based near-field optical measurements are required. We have further developed an ultrathin-film encapsulated technology for K-SiN-Air device in non-vacuum conditions (see fabrication details in Supporting Information S. II. 11), based on which the near-field optical spectroscopy such as the scattering-type scanning near-field optical microscope (s-SNOM)[39] is available for probing the dispersion relation characteristics of excited SPP mode [Fig. 4(a)]. The s-SNOM imaging data were collected at incident laser wavelength of 700 - 900 nm [Fig. 4(b)]. Fig. S17(a) plots the near-field fringe profiles and the corresponding FT profiles are shown in Fig. S17(b), which is expected as

$$\frac{k_p}{k_0} = \frac{2\pi/\rho}{k_0} + cos\alpha, \qquad (3)$$

where $\alpha = 36°$ is the incident angle of laser beam relative to sample plane. Fig. 4(c) displays the obtained ($k_p/k_0$, $E$) data points, illustrating the dispersion relation of the SPP mode, which exhibits good agreement with the calculated dispersion results (see calculation details in Supporting Information S.I.4). The SPP effective index $k_p/k_0$ for K exceeds 10, which is almost one order of magnitude than the mainstream low-loss counterparts of silver and emerging transparent conductor TMTTF)$_2$SbF$_6$. Note that the latter can only be metallic for $\lambda > 1900$ nm (see Fig. 3), its optical concentration depicted in Fig. 4b is merely due to the higher refractive index of bulk TMTTF)$_2$SbF$_6$. One can further promote the available optical confinement of K towards the theoretical prediction (red solid line) provided that Landau damping can be effectively reduced, which dominates the loss mechanism when the field is concentrated in the narrow surface layer at extreme large wave-vectors[27]. Furthermore, other geometries, such as gap SPP or hyperbolic configurations, can provide extra feasibility to attain even greater confinement. That said, despite the limitations imposed by Landau damping, the deep-subwavelength optical confinement results obtained here further underscore the advantages of potassium as a plasmonic material, which exhibits exceptionally low intrinsic loss.

**Conclusion**

In this work, we have demonstrated, both theoretically and experimentally, the potential of potassium as a plasmonic metal exhibiting record-low optical loss over a broad spectral range, enabled by the SOC fabrication process. Owing to the suppression of U-scattering-related optical absorption and the achievement of oxide-free crystallization, the SOC-grown potassium films exhibit an exceptionally low imaginary part of the permittivity (~ 0.1) across the entire visible and near-infrared regions, corresponding to a minimal damping rate of 3.7 meV—over 1 - 2 orders of magnitude lower than that of noble metals and/or even transparent conductors. This breakthrough further enables K-based SPP devices to achieve unprecedented deep subwavelength optical confinement, with a compression factor approaching 10 in an unpatterned configuration. Overall, our findings establish a viable pathway toward realizing the ultimate low-loss limit of plasmonic metals and open new avenues for advancing deep subwavelength photonics.

**Method**

**FIB fabrication.** We use gallium ion equipped double beam electron microscope (Scios DualBeam from FEI, 30 eV) to cut cross section of the potassium film. The active potassium film sample was sent into the electron microscope cavity with a vacuum box to prevent oxidation by air (Quick Loader IGST SDB from FEI).

Acknowledgments

We acknowledge the micro-fabrication center of National Laboratory of Solid State Microstructures (NLSSM) for technique supports. This work was supported by the National Key Research and Development Program of China (2021YFA1400700, 2022YFA1404300); the National Natural Science Foundation of China (12022403, 62375143) and the Natural Science Foundation of Jiangsu Province (BK20243009).

Author contributions:

L.Z. proposed the research. J.Z. and S.Z. guided the project. Y.Y., P.Y. and H.F. designed the experiments. Y.Y., Y.Z., P.Y. and Y.S. contributed equally to this work. Y.Y., P.Y. and H.F. finished the sample fabrication, spectroscopic ellipsometer test, XRD test, numerical simulation and data analysis. J. K. guided the theoretical part and Y.Z finished the theory calculation. Y.Y. and Y.S. finished the s-SNOMs test. Y.Y. and P.Y. finished the data presentation. Y.Y., Y.Z, J.K. and L.Z. wrote the paper with input from all authors.

Competing interests:

Authors declare that they have no competing interests.

Data availability:

We declare that the data supporting the findings of this study are available within the paper.

Additional information:

Supporting information

Correspondence and requests should be addressed to Lin Zhou.

Supplementary movie S1

Supplementary Materials for

# Approaching the Low Optical Loss Limit of Plasmonics using Potassium Metals

Yuhan Yang[1]†, Yi Zhang[2]†, Yuhong Shi[1]†, PengCheng Yao[1]†, Hanyu Fu[1], Jacob B. Khurgin[3]*, Fengrui Hu[1], Jia Zhu[1], Shining Zhu[1], Lin Zhou[1]*

*Corresponding author: linzhou@nju.edu.cn; jakek@jhu.edu

## Contents

# S. I Calculation and theory details

## S. I. 1 The calculation of intraband loss induced by three kinds of scattering

The expression for the normal e-p scattering at room temperature is given by[1]

$$v_{ep}{}^N = \frac{\pi E_d{}^2 n_F}{2\hbar \rho s^2} k_B T$$

(S1)

where $E_d$ is deformation potential, $n_f$ is density of states at $E_F$, $\rho$ is density of materials and $s$ is velocity of sound in the material. For free electron model, the deformation potential is equal to $-2\,E_F/3$. Substituting the result into Eq.(S1), one can obtain the Eq.(2) in main text. However, for the umklapp e-p scattering, the rate of e-p scattering can be written as[2]

$$v_{ep}{}^U = \alpha v_{ep}{}^N \quad (S2)$$

where $\alpha = \frac{v_1}{v_2}\left(\frac{E_{d,t}}{E_{d,l}}\right)^2 \left(\frac{s_l}{s_t}\right)^4 I(x_0)$, $v_1$ and $v_2$ are the electron velocities on the two fermi surfaces located in two adjacent Brillouin zone, respectively, and the $E_{d,t}$ and $E_{d,l}$ are longitudinal and transverse components of the deformation potential responsible for scattering by the longitudinal and transverse phonons with velocities $s_t$ and $s_l$, and $I(x_0) = \int_{x_0}^{\infty} x^2 \Phi(x) dx$, where $x_0 = \hbar q_{min} s / k_B T$ (here $s$ is the average velocity of sound in the material and $q_{min}$ is the smallest gap between the two fermi surface in two adjacent Brillouin zone) and $\Phi(x) = \frac{4}{(e^x-1)(e^{-x}+1)}$ is occupation factor at fermi level. The method of calculation of $E_{d,t(l)}$ can be found in ref.[3] and ref.[4].

Several works[5,6] made evident that the umklapp scattering of electrons is major source of the optical loss of metals at low temperature. Therefore, that means the umklapp scattering is the main mechanism of optical loss in metals induced by e-e

scattering. The e-e scattering rate can be written as:

$$\gamma_{e-e} = \frac{\pi^3 \Gamma \Delta}{12 \epsilon_F}\left[(k_B T)^2 + \left(\frac{\hbar\omega}{2\pi}\right)^2\right], \tag{S3}$$

where $\Gamma$, $\Delta$, $\epsilon_F$ and $k_B$ are the average scattering probability, fractional umklapp scattering, Fermi energy and Boltzmann constant, respectively. The methods of calculation of $\Gamma$ and $\Delta$ are given in Ref. [5].

The e-s scattering can be calculated by the formula[7]

$$\gamma_{es} = \frac{3}{8}\frac{v_F}{\delta} \tag{S4}$$

where $v_F$ is Fermi velocity and $\delta$ is the skin depth. The loss due to the e-s scattering is, in fact, the Landau damping or Kreibig damping. Due to combined lower Fermi velocity and large skin depth ($\delta = \sqrt{2/\omega\mu\sigma}$, where $\omega$ is the frequency of incident light, $\mu$ is permeability and $\sigma$ is conductivity) compared to other metals, alkali metals exhibit extremely low e-s scattering rate. As an example, based on the corresponding data of $v_F$ and $\delta$, the electron-surface scattering rate of potassium is about $0.81 \times 10^{13} s^{-1}$, while that of silver and gold are $4.17 \times 10^{13} s^{-1}$ and $5.25 \times 10^{13} s^{-1}$, respectively.

## S. I. 2 First ionization energy of alkali metals

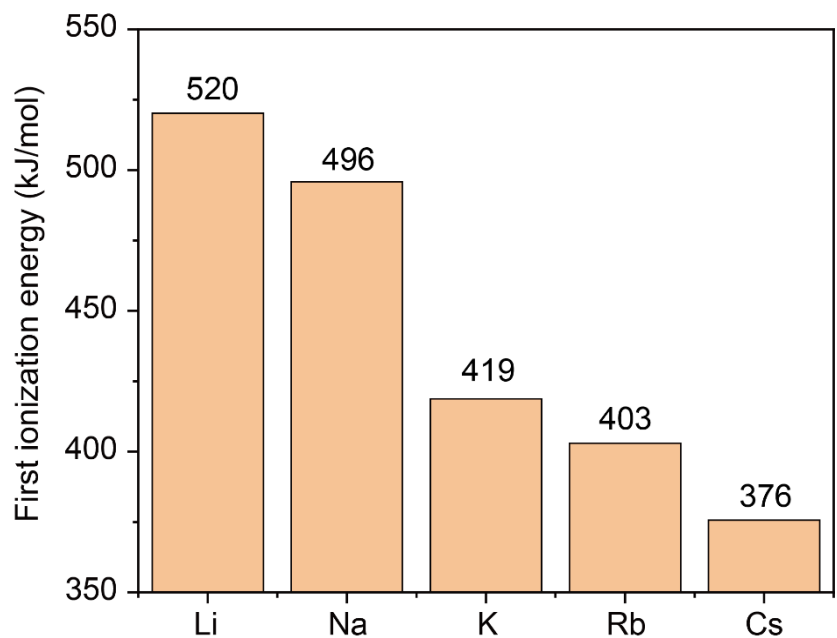

**Figure S3.** First ionization energy of alkali metals. The first ionization energy of K relative to Na is reduced by 15%, which is higher than that of other metals in the same group (~ 5%).

**S. I. 3 DFT calculations of binding energy**

In order to figure out the mechanism of quartz peeling process, we made DFT calculations of the binding energy between K$_2$O/K pair and K$_2$O/SiO$_2$ using the CASTEP code integrated into the Materials Studio software. Ultrasoft pseudopotential as implemented in the CASTEP code was applied. PBE form of generalized-gradient approximation (GGA) was adopted. The K/K$_2$O pair was confined in a 5.32*7.52*43.34 Å$^3$ periodic box, with the K (1,1,0) surface and K$_2$O (1,1,0) surface initially exposed. The K/SiO$_2$ pair was confined in a 5.05*5.05*34.94 Å$^3$ periodic box, with the SiO$_2$ (0,0,1) surface and K$_2$O (1,1,0) surface initially exposed. K, K$_2$O, and SiO$_2$ alone, as well as the K/K$_2$O pair and K/SiO$_2$ pair all underwent geometry optimization, respectively, without specific lattice symmetry restrictions to simulate the amorphous state of each material. The geometry optimization was done by means of the Broyden-Fletcher-Goldfarb-Shanno (BFGS) algorithm, allowing all atomic dimensions to vary to reach the minimum energy state. The binding energy of materials A and B was calculated from the equation below:

$$E_{binding} = E_A^{total} + E_B^{total} - E_{A/B}^{total} \qquad (S12)$$

The optimized superlattice structures and binding energies per unit area are shown in **Figure S4**. The result indicates that K/K$_2$O pair has a lower binding energy than K/SiO$_2$ pair, which enables the of K$_2$O attach to SiO2 substrate at initial point.

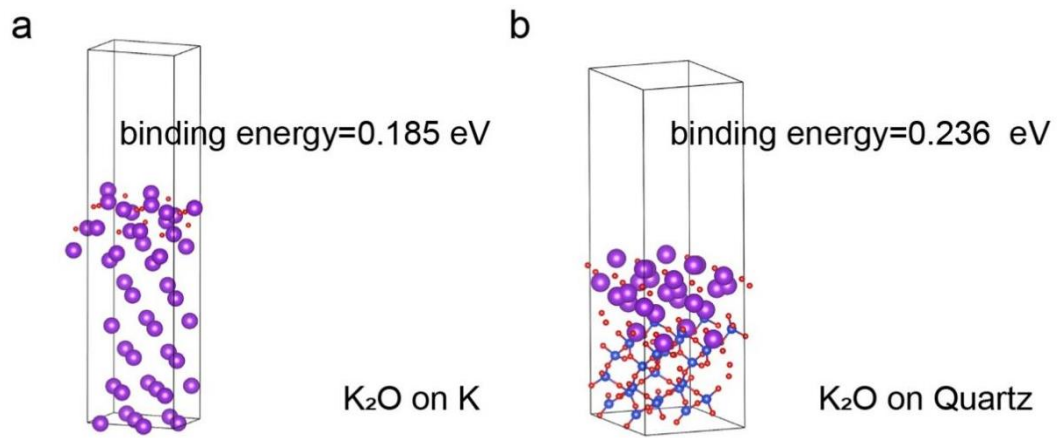

**Figure S4.** Optimized atomic structure of $K_2O$ (a) and K (b) on quartz. The binding energy between $K_2O$ and K ($E_{K_2O/K}$) is lower than that between $SiO_2$ and K ($E_{K_2O/SiO_2}$).

## S. I. 4 Calculation of dispersion curve of multilayer Air/SiN/Metal devices

The dispersion curve of our multilayer K-SiN-Air devices in **Figure 4e** was obtained from analytical solution derived from Maxwell's equations. Considering the electric field right underneath the s-SNOM probe is perpendicular to the sample surface, only transverse magnetic (TM) waveguide modes are excited.

We utilized a multilayer planar waveguide structure to calculate the analytical solutions of the SPP modes excited in the Metal-SiN-Air devices. Due to the trace amounts residual oxygen at scattering edges observed in experiment, which tend to interact with the reactive metal (especially potassium) at the interface of metal and SiN, we consider an additional oxide layer ($\varepsilon_2 = 2.25$) between the SiN ($\varepsilon_1 = 4.62$) and metal layer as a correction layer. The structure is shown in **Figure S5**. The dispersion equation for the SPP modes excited within the multilayer structure is derived using Maxwell's equations and boundary conditions, as follows:

$$e^{-2k_2 b} = \frac{Y_1 \left(\frac{k_4}{\varepsilon_4} + \frac{k_2}{\varepsilon_2}\right)}{Y_2 \left(\frac{k_4}{\varepsilon_4} - \frac{k_2}{\varepsilon_2}\right)} \tag{S13}$$

$$Y_1 = \frac{2k_1}{\varepsilon_1}\left(\frac{k_3}{\varepsilon_3} + \frac{k_1}{\varepsilon_1}\right)e^{2k_1 a - k_2 a} - \left(\frac{k_2}{\varepsilon_2} - \frac{k_1}{\varepsilon_1}\right)e^{-k_2 a}\left[\left(\frac{k_3}{\varepsilon_3} - \frac{k_1}{\varepsilon_1}\right) - \left(\frac{k_3}{\varepsilon_3} + \frac{k_1}{\varepsilon_1}\right)e^{2k_1 a}\right] \tag{S14}$$

$$Y_2 = \frac{2k_1}{\varepsilon_1}\left(\frac{k_3}{\varepsilon_3} + \frac{k_1}{\varepsilon_1}\right)e^{2k_1 a + k_2 a} + \left(\frac{k_2}{\varepsilon_2} + \frac{k_1}{\varepsilon_1}\right)e^{k_2 a}\left[\left(\frac{k_3}{\varepsilon_3} - \frac{k_1}{\varepsilon_1}\right) - \left(\frac{k_3}{\varepsilon_3} + \frac{k_1}{\varepsilon_1}\right)e^{2k_1 a}\right] \tag{S15}$$

$$k_1 = \sqrt{k_p^2 - k_0^2 \varepsilon_1} \tag{S16}$$

$$k_2 = \sqrt{k_p^2 - k_0^2 \varepsilon_2} \tag{S17}$$

$$k_3 = \sqrt{k_p^2 - k_0^2 \varepsilon_3} \tag{S18}$$

$$k_4 = \sqrt{k_p^2 - k_0^2 \varepsilon_4} \tag{S19}$$

Here, $\varepsilon$ represents the dielectric constant of the respective layer, $a = 40$ nm is the thickness of the SiN layer, set to 40 nm, and $b$ is the thickness of the oxide correction layer. For the K system, $b$ is set to 3 nm; for the Na system, it is set to 2 nm; and for the Ag system, it is set to 0 nm. The thickness $b$ is proportional to the chemical reactivity of the three metals.

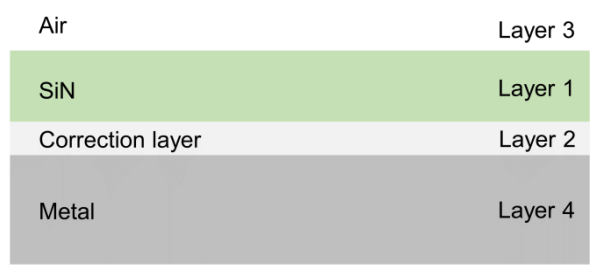

**Figure S5.** The illustration of the Air/SiN/Correction layer/Metal multilayer planar structure.

# S. II Experiment details

**S. II. 1 Potassium film fabricated by the thermal-assisted spin-coating method:**

We have tried employing the conventional thermal-assisted spin-coating method[8] to fabricate potassium films (**Figure S6a**). However, potassium's greater reactivity with oxygen than sodium leads to the newly formed oxide nearly enveloping the target substrate's surface during fabrication (**Figure S6b**). Additionally, optical microscopy uncovers randomly distributed impurities (**Figure S6c**). **Figure S7** presents the ellipsometry data for K films fabricated using the thermal-assisted spin-coating method, with a comparison to Na films. We didn't observe the intrinsic superior optical loss of K through the imaginary parts of the dielectric functions.

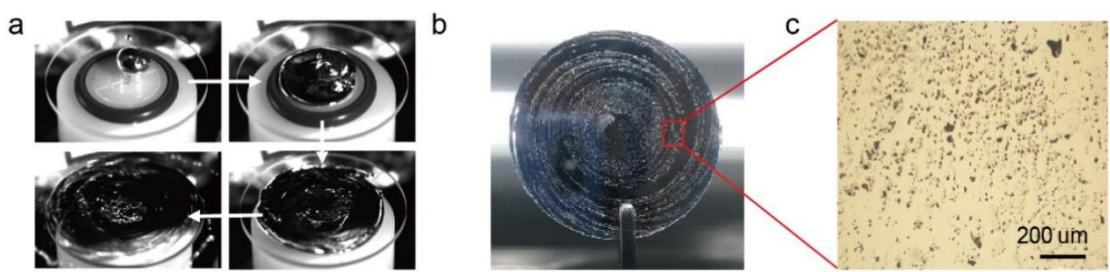

**Figure S6.** a) experimental process of the thermal-assisted spin-coating[8]. b) photograph of potassium film fabricated by thermal-assisted spin-coating method. c) partial magnification of the light microscope image of the potassium film.

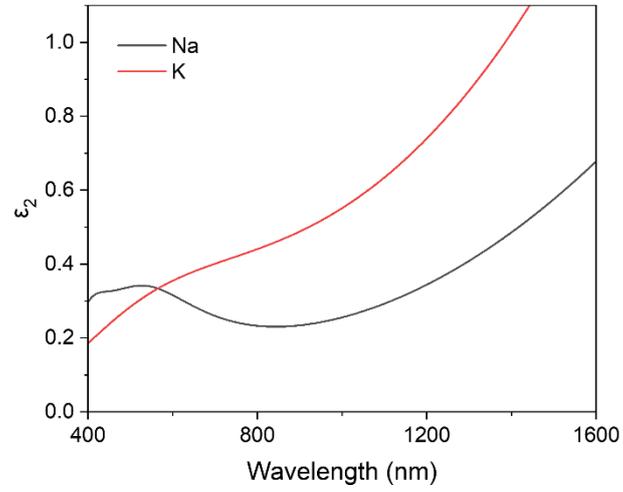

**Figure S7.** Comparison of imaginary parts of the dielectric functions of potassium film and sodium film[8] fabricated by thermal-assisted spin-coating method, measured with the spectroscopic ellipsometer.

## S. II. 2 The oxide removal process

By heating the quartz tube to ~ 120 °C (above potassium's melting point of ~ 64 °C), we maintained the internal potassium in a liquid state. Then, we squeezed the latex at tube's head to keep the liquid K flow through the quartz tube. This process effectively removed the original oxide layer, leaving an oxide shell on the quartz tube's inner surface. Enlargement of **Figure S8a** shows the schematic and optical photograph of the process, with purified liquid on the topside of the tube and the remaining oxide trail on the underside. XRD analysis of the residual white layer on the quartz sidewall in **Figure S8b** confirmed it as $K_2O$. Based on this process, the fresh potassium flow is available from the spray head of the heated quartz tube for further film preparation.

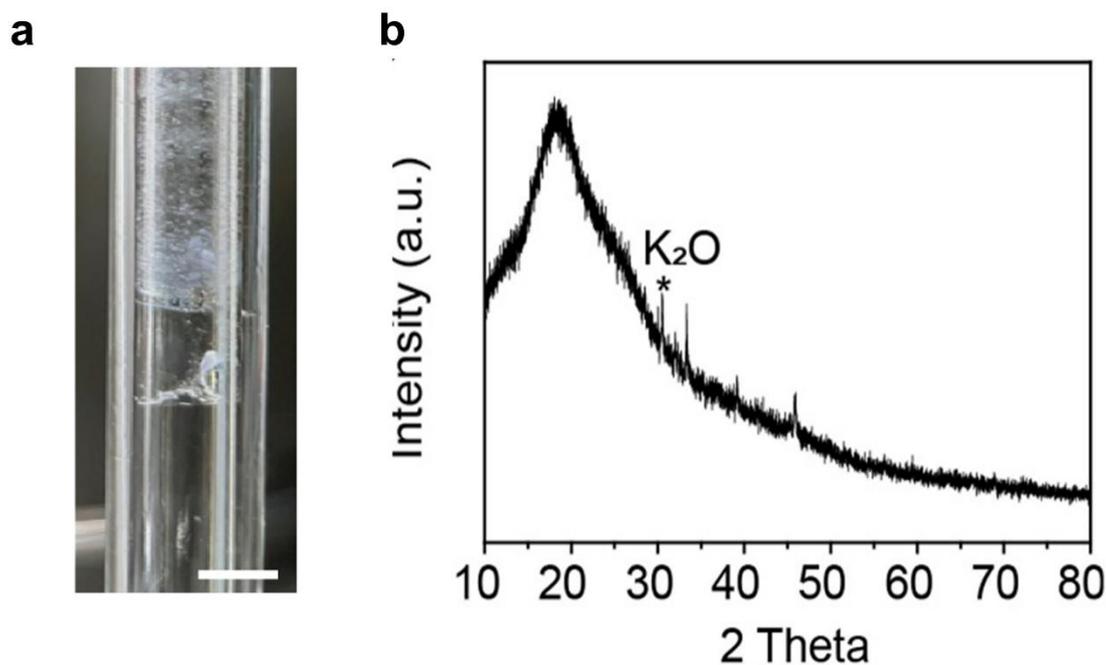

**Figure S8.** (a) The optical paragraph of oxide trail left on quartz tube. (b) XRD test result of the oxide shell on the trail after the liquid potassium flows through.

### S. II. 3 The crystallization process

Once the fresh potassium is produced, the key is to form the high-crystal quality potassium before further oxidation. As illustrated in **Figure 2c-e**, the droplet undergoes a free-fall and slips against a rapidly moving ultra-flat quartz substrate (act as the sharp blade). While it is still possible to form a thin potassium oxide during this period, the large tangential-velocity mismatch between potassium liquid (with a thin oxide) and quartz substrate induces huge lateral adhesion forces ($F_{AF}$) larger than the horizontal component of surface tension ($\gamma_x$), allowing fresh liquid potassium to flow out and directly adhere to the substrate without oxygen exposure (**Figure 2c**). The potassium then solidifies along the path, forming a high-quality K film (**Figure 2d**) with an oxide-free interface at the substrate.

More specifically, when K droplet slides on an ultra-flat stiff surface, the lateral adhesion forces ($F_{AF}$) can be expressed as[9]:

$$F_{AF}(v) = \frac{24}{\pi^3}\gamma D\bigl(\cos\theta_r(v) - \cos\theta_a(v)\bigr), \tag{2}$$

where $\gamma$ represents the surface tension coefficient, 106 mN/m[10] for potassium liquid. $D$ denotes the diameter of the contact region. $\theta_a$ and $\theta_r$ stands for the dynamic advancing and receding contact angles and are highlighted in **Figure S10**, respectively. These two contact angles depend on the relative velocity $v$ of the droplet with respect to substrate: as $v$ increases, $\theta_a$ increases, whereas $\theta_r$ decreases. It thus can be inferred that $F_{AF} \propto v$. In other words, for the process of **Figure 2c**, when $v$ is sufficiently high, $F_{AF}$ can exceed the surface tension of the potassium droplet, triggering a crack in droplet's surface and imitating a rupture in the oxide shell.

Therefore, it is important that the potassium slips on a high-speed substrate. It is observed that in our optimized experimental condition with $v$ reaching 12 m/s, the contact angles between the droplet and substrate can be measured in situ with the high-speed camera (**Figure S10**). $\theta_a$ is about 113° and $\theta_r$ is about 9°. It can be calculated that $F_{AF}$ reaches approximately 4.5 mN, significantly surpassing the droplet's horizontal component of surface tension of 1.1 mN*cos (67°)=0.43 mN (characteristic droplet size ~1 cm), leading to the split of the potassium drop's surface and thus ensuring the instantaneous homogeneous potassium film formation with oxidation effectively excluded.

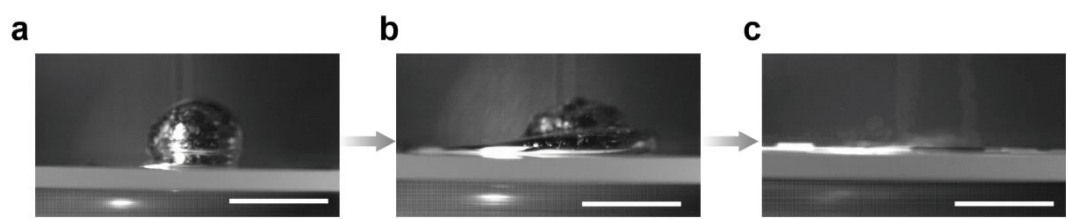

**Figure S9.** Optical images corresponding to the three main processes of crystallization process of SOC method, which taken by high-speed camera (scale bar: 5 mm).

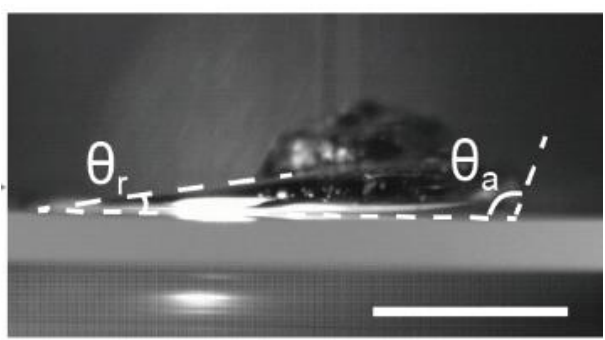

**Figure S10.** The highlighted $\theta_a$ and $\theta_r$ of slipping K droplet. Scale bar: 5 mm.

## S. II. 4 Preparation process of ultra-smooth potassium samples for XRD tests

The ultra-smooth potassium film samples for XRD tests is sealed with polyimide tape, an ionomer resin transparent to X-rays for a wide range of angles, and the process is shown in **Figure S11**.

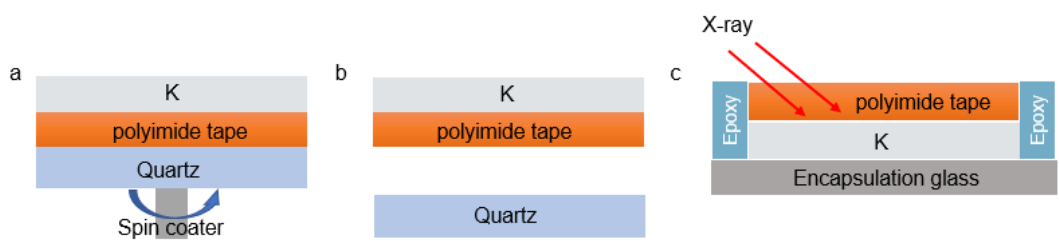

**Figure S11.** Preparation process of ultra-smooth potassium samples for XRD tests.

## S. II. 5 Crystalline nature comparation of potassium film and sodium film fabricated by different methods

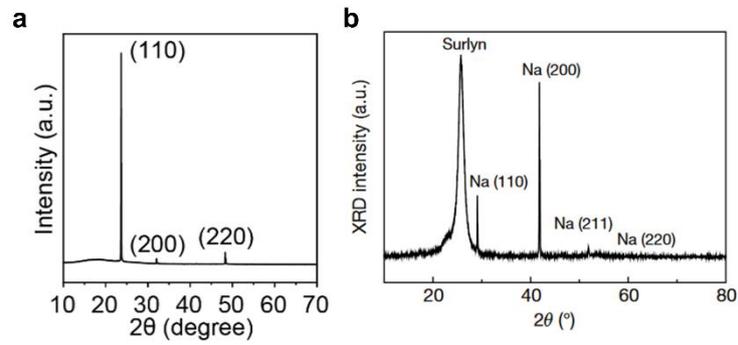

**Figure S12.** XRD test result of prepared K (a) and Na (b) film fabricated by different methods.

**S. II. 6 Analysis of the optical image of the prepared potassium film**

**Figure S12** illustrates a typical K film prepared using our SOC method. The entire trajectory traversed by the K droplet is solidified into a smooth and uniform K film, corresponding to the process shown in **Figure 2c-e**. Notably, the white mark at the initial point represents an oxidation layer left by the initial contact between the freely falling K droplet and the substrate. During the subsequent slipping process, the liquid K directly contacts the substrate and solidifies into the film, without any interfacial inclusion of the oxide layer.

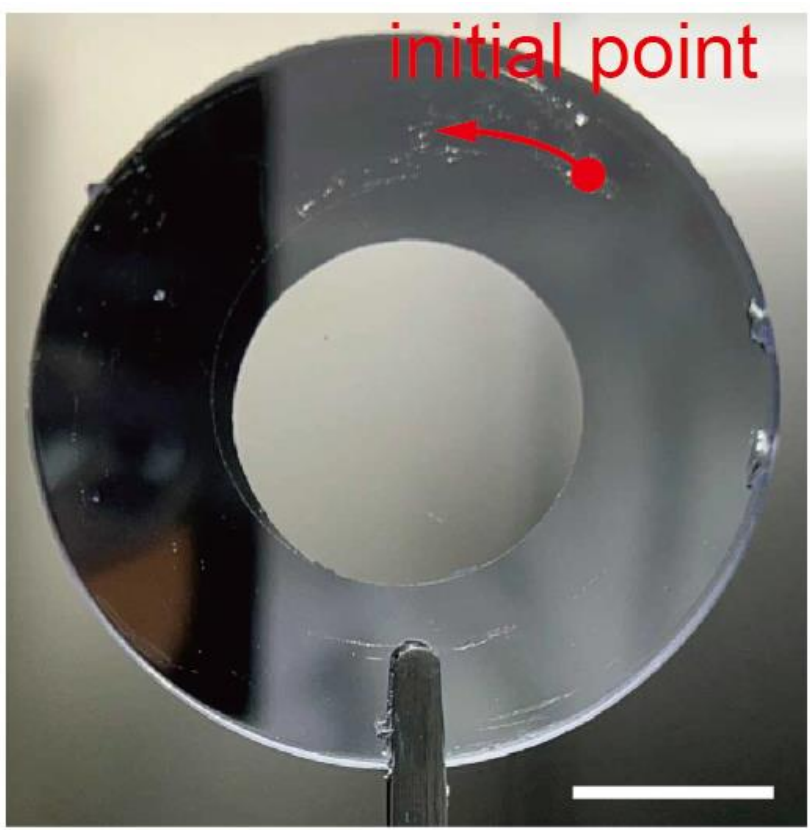

**Figure S13.** A typical K film prepared using the SOC method. The whole solidified potassium film along slipping trail is smooth while the initial point remains several oxidations (scale bar: 1cm).

## S. II. 7 Structure of semi-infinite SPP waveguide for measuring propagation length

We fabricated potassium-based plasmonic waveguides in near-infrared to fit the optical loss from propagation length $L_{spp}$. The SPP mode in waveguides were excited through gratings. By varying period of grating ($P$), we controlled the excitation wavelength of SPP, while keeping the fill factor at approximately 0.5 and the grating depth at around 150 nm. **Figure S13** shows the schematic (upper panel) and experimental (lower panel) configurations of the potassium-based plasmonic waveguide device, respectively. The left coupler (noted as downward red arrow) converts the incident laser beam to SPPs (left light spot), which propagate along the potassium–quartz interface and then out-coupled to free space via the right coupler (right light spot).

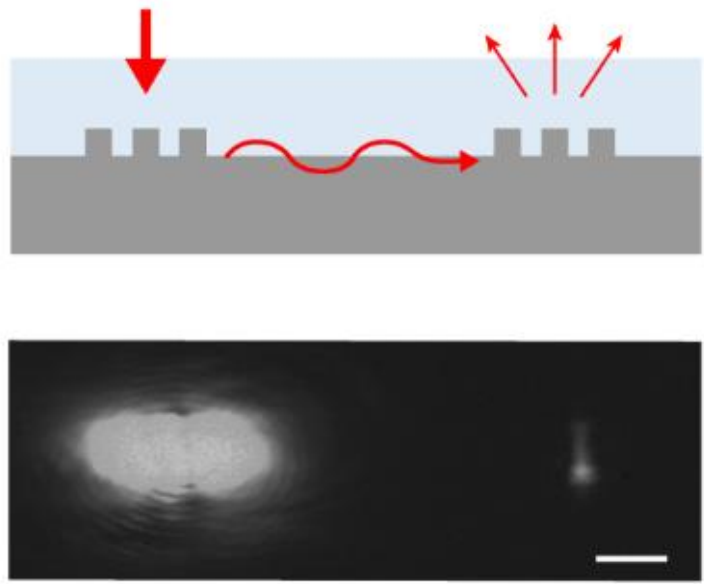

**Figure S14.** Schematic of potassium-based plasmonic waveguiding structure for propagation length measurement (up) and related optical image of the light spots (bottom). The scale bar is 20 μm.

**S. II. 8 The γ retrieved from propagation length and compared with sodium from previous literature**

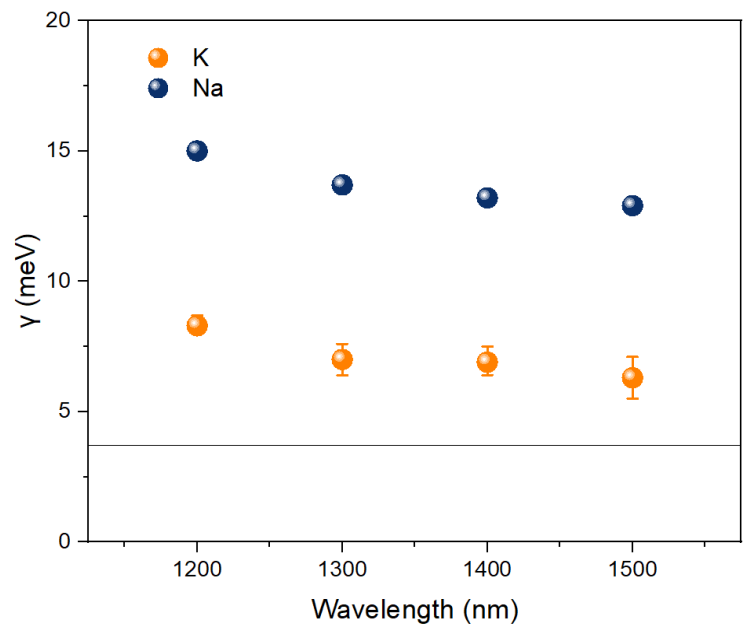

**Figure S15.** The magnified of γ retrieved from propagation length and compared with sodium from previous literature with error bar.

**S. II. 9 Propagation measurements for potassium film with exponential curves fitted to the data**

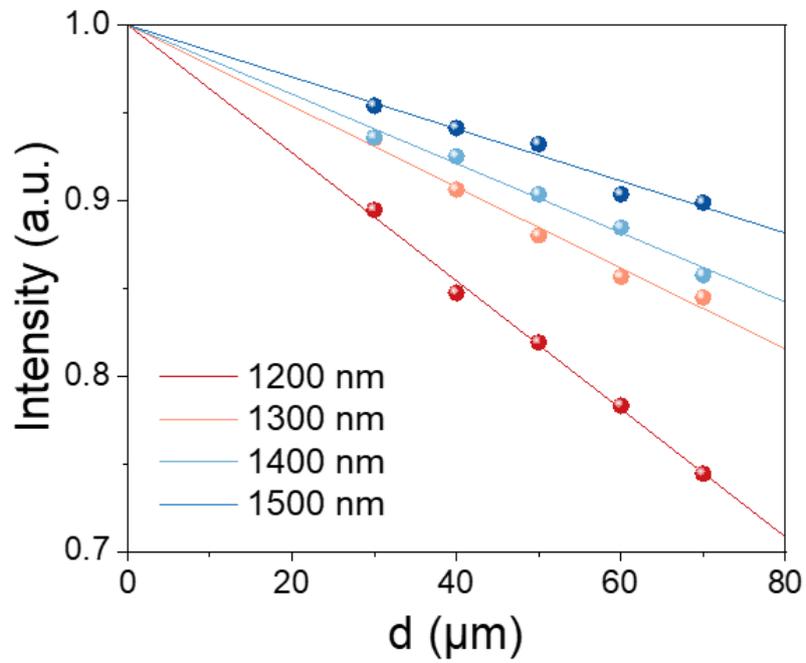

**Figure S16.** Propagation measurements at wavelengths of 1200 nm, 1300 nm, 1400 nm and 1500 nm for potassium film, with exponential curves fitted to the data.

**S. II. 10 Optical loss extracted from propagation length through plasmonic waveguide device**

As the intensity of SPP decreases exponentially as a function of distance, we fitted the propagation length from various propagation separations with different distances as follow:

$$lgI = -\frac{1}{L_{spp}}d + lgI_0, \tag{S20}$$

where $I$ is the intensity of the outgoing light, $I_0$ is the intensity of the incident light and $d$ is the distance between the two couplers. We then extracted optical loss $\gamma$ from propagation length as follow:

$$L_{spp} = \frac{\lambda}{2\pi\varepsilon_R\varepsilon_I}\left[\frac{\varepsilon_R(\varepsilon_R+\varepsilon_I)}{\varepsilon_d}\right]^{\frac{3}{2}}, \tag{S21}$$

$$\varepsilon_R = 1 - \frac{\omega_p^2}{\omega^2+\gamma^2}, \tag{S22}$$

$$\varepsilon_I = \frac{\omega_p^2\gamma}{\omega(\omega^2+\gamma^2)}, \tag{S23}$$

where $\varepsilon_R$ and $\varepsilon_I$ are the real and imaginary parts of the metal permittivity respectively, $\varepsilon_d$ is the dielectric constant of the quartz substrate, which is taken as 1.46, $\gamma$ is the fitted optical loss of potassium and the plasma frequency $\omega_p$ is generally related to the carrier concentration in the metal and is set to 4.0 eV.

**S. II. 11 Fabrication of metal/SiN/Air SPP device for s-SNOMs test**

We purchased TEM SiN grids (AR010C) with the grid size of 100 μm and the covered SiN thickness of ~ 40 nm. Using the TEM SiN grid as a substrate, we fabricated the K or Na film via the SOC process and Ag film via the physical vapor deposition (PVD) process, forming a metal/SiN/Air structure at the window. The 40 nm thick SiN layer ensures that oxygen do not come into contact with the metal film, while also providing a measurement window for s-SNOM testing.

**S. II. 12 Real-space fringe profiles and corresponding FT profiles of the s-SNOM imaging data**

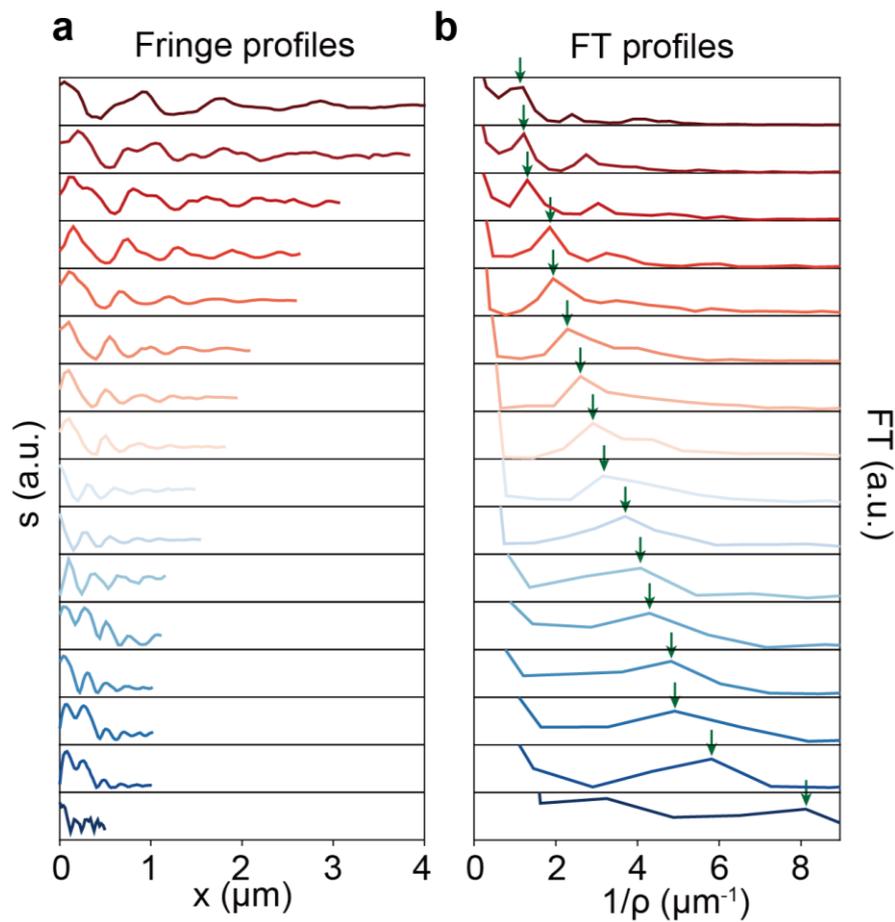

**Figure S17.** Real-space fringe profiles (a) and corresponding FT profiles (b) of the s-SNOM imaging data of Fig. 4(b).